\documentclass[aps, prb, preprint, superscriptaddress,preprintnumbers,amssymb]{revtex4-2}
\usepackage{graphicx}
\usepackage{dcolumn}
\usepackage{bm}
\usepackage{color}

\begin{document}

\preprint{v7.1}

\title{Topological frequency shift of quantum oscillation in CaFeAsF}



\author{Taichi Terashima}
\email{TERASHIMA.Taichi@nims.go.jp}
\affiliation{International Center for Materials Nanoarchitectonics, National Institute for Materials Science, Tsukuba 305-0003, Japan}
\author{Shinya Uji}
\affiliation{International Center for Materials Nanoarchitectonics, National Institute for Materials Science, Tsukuba 305-0003, Japan}
\author{Teng Wang}
\author{Gang Mu}
\email{mugang@mail.sim.ac.cn}
\affiliation{State Key Laboratory of Functional Materials for Informatics, Shanghai Institute of Microsystem and Information Technology, Chinese Academy of Sciences, Shanghai 200050, China}
\affiliation{CAS Center for Excellence in Superconducting Electronics (CENSE), Shanghai 200050, China}


\date{\today}


\maketitle

Guo, Alexandradinata, \textit{et al.} have recently proposed that quantum-oscillation frequencies from Dirac/Weyl fermions exhibit a negative shift proportional to $T^2$ because of the energy dependence of the effective mass peculiar to a linear band-dispersion.
We have measured Shubnikov--de Haas oscillation in CaFeAsF up to $T$ = 9 K.
The frequency of the $\alpha$ Dirac electron exhibits a negative shift with increasing $T$, while that of the $\beta$ Schr\"odinger hole does not.
For $T \geqslant 5$ K where $\beta$ is negligible, the $\alpha$-frequency shift is proportional to $T^2$ and its rate agrees with the theoretical prediction within experimental accuracy.
At lower temperatures, the shifts of $\alpha$ and $\beta$ deviate from theoretical expectations, which we ascribe to the inaccuracy in the frequency determination due to unfavorable interference between frequencies.
Our results confirm that the topological frequency shift can be utilized to identify Dirac/Weyl fermions when quantum-oscillation frequencies can be determined accurately.

\newpage
\section*{Introduction}
Quantum oscillation arising from Landau quantization of electron motion in magnetic fields is a powerful tool to investigate electronic structures of metals.
It dates back to 1930 when quantum oscillation in the electrical resistance and magnetization in Bi was observed \cite{SdH, dHvA}.
Since the Onsager relation \cite{Onsager52PM} and Lifshitz-Kosevich formula \cite{Lifshitz52PM} to interpret the quantum oscillation were established in 1950s, it has been used to determine the Fermi surface of not only elemental metals \cite{Shoenberg84} but also heavy fermions \cite{Reinders86PRL, Taillefer88PRL}, high-$T_{\mathrm{c}}$ cuprates \cite{Doiron-Leyraud07Nature}, and so on.
The latest field of its application is topological materials.
Topological surface states of topological insulators were successfully observed in quantum-oscillation measurements \cite{Qu10Science,Analytis10NatPhys}.
Further, Dirac or Weyl fermions in topological semimetals can in principle be identified by detecting the Berry phase in quantum-oscillation measurements \cite{Mikitik99PRL, Mikitik98JETP}.
However, the necessary procedure is involved even if it is not impossible to perform.

To be specific, we focus on Shubunikov-de Haas (SdH) oscillation, which is described as follows \cite{Richards73PRB, Shoenberg84}:
\begin{equation}\label{LK}
\frac{\Delta\rho}{\rho_0} = -C\sum_{r=1}^{\infty} \sqrt{B}R_{\mathrm{T},r} R_{\mathrm{D},r} R_{\mathrm{s},r} \cos\left[2\pi r \left(\frac{F}{B}-\frac{1}{2}\right)+ \phi_{\mathrm{D}}+r \phi_{\mathrm{B}}\right],
\end{equation}
where $\Delta\rho$ is the oscillatory part of resistivity, while $\rho_0$ is the background resistivity.
For simplicity, we have assumed $\rho = \sigma^{-1}$ neglecting the tensorial nature of $\rho$ and $\sigma$.
$C$ is a positive coefficient.
The frequency $F$ is related to a Fermi-surface cross sectional area $A$ as $F = (\hbar/2\pi e)A$.
The SdH oscillation contains not only the fundamental frequency $F$ but also its harmonics ($r > 1$).
The temperature and Dingle reduction factors are given by $R_{\mathrm{T},r} = rX/\sinh(rX)$ and $R_{\mathrm{D},r}=\exp(-rX_{\mathrm{D}})$, where $X_{(\mathrm{D})}=K\mu^* T_{(\mathrm{D})}/B$, $\mu^*=m^*/m_{\mathrm{e}}$, and the coefficient $K$ is 14.69 T/K.
The spin reduction factor $R_{\mathrm{s},r}$ describes the interference of oscillations from up- and down-spin electrons and is given by $R_{\mathrm{s},r}=\cos (r \pi S)$, where the spin-splitting parameter $S$ is given by $g \mu^*/2$, $g$ being the conduction-electron $g$ factor averaged over the cyclotron orbit under consideration.
$\phi_{\mathrm{D}}$ is 0 for a two-dimensional (2D) Fermi-surface (FS) cylinder while it is + or $-\pi/4$ when the oscillation is from a minimum or maximum cross section of a three-dimensional (3D) FS pocket.
$\phi_{\mathrm{B}}$ is the Berry phase, which is 0 for normal (i.e., Schr\"odinger) fermions but $\pi$ for Dirac fermions \cite{Mikitik99PRL, Mikitik98JETP}.

We note that the phase of fundamental quantum oscillation is the same between normal fermions with a negative $R_{\mathrm{s},1}$ and Dirac fermions with a positive $R_{\mathrm{s},1}$.
Therefore, if we like to determine the Berry phase from experimental data by making a Landau-index plot or fitting Eq. 1 to the data, we first have to determine the sign of $R_{\mathrm{s},1}$.
$R_{\mathrm{s},1}$ depends on $g$ and $m^*$.
$g$ may substantially deviate from the free-electron value $g$ = 2, especially for small orbits: e.g. for the Zn needle, $g >$ 100 is generally accepted \cite{Shoenberg84}.
$g$ could be determined by e.g. the spin-zero method in some special cases \cite{Terashima18PRX}, but it is impossible in many cases.
Note that even in noncentrosymmetric crystals, in which double degeneracy of electronic bands is lifted, $R_{\mathrm{s},r}$ (or its equivalent) has to be considered as long as time-reversal symmetry is preserved because oscillations from a time-reversal pair of orbits interfere \cite{Hirose20PRB}.

Recently, Guo, Alexandradinata, \textit{et al.} (hereafter GAM \textit{et al.} after the three corresponding authors) proposed a new approach \cite{Guo21NatCommun}.
They pointed out that quantum-oscillation frequency from Dirac/Weyl pockets should exhibit a characteristic temperature dependence.
The cyclotron effective mass $m^*$ is defined as $m^* = (\hbar^2/(2\pi)) |\partial A/\partial E|$.
Accordingly, $\partial m^* / \partial E$ is zero for a quadratic band dispersion but finite for a linear dispersion:
assuming $E = \hbar v k$, where $v$ is the Fermi velocity, $m^* = |E|/v^2$ and $|\partial m^* / \partial E| = 1/v^2$, that is, as $|E|$ decreases both $A$ and $m^*$ decrease. 
At a finite temperature $T$, the Fermi edge broadens, and hence the system might be viewed as a sum of hypothetical systems with the Fermi energy distributed over a range of $\sim k_{\mathrm{B}} T$ around $E_{\mathrm{F}}$.
Since oscillation from smaller orbits with smaller $m^*$ survives to higher temperatures, the effective frequency might decrease.
They extended the Lifshitz-Kosevich formula to the next order in $k_{\mathrm{B}} T / E_{\mathrm{F}}$ and showed that the above expectation is indeed the case (see also \cite{Fortin15EPJB}):
\begin{equation}\label{GAM}
F(T) = F_0 -\frac{(\pi k_{\mathrm{B}} T)^2}{4\beta}\frac{1}{m^*} \left| \frac{\partial m^*}{\partial E} \right|,
\end{equation}
where $F_0$ is the frequency at $T$ = 0 and $\beta = e \hbar/(2m^*)$.
They analyzed quantum oscillation from Dirac pockets in Cd$_3$As$_2$ and LaRhIn$_5$ and demonstrated that Eq. 2 combined with another small correction to $F(T) $, which will be described later, could explain the experimentally observed temperature dependence of the frequencies.

In this article, we apply the GAM method to SdH oscillation in CaFeAsF, an iron-based superconductor parent compound.
The Fermi surface in CaFeAsF is composed of a pair of $\alpha$ Dirac electron cylinders and a $\beta$ Shr\"odinger hole cylinder \cite{Terashima18PRX}.
By virtue of the quasi-two-dimensional electronic structure, the sign of $R_{\mathrm{s},1}$ was unambiguously determined for both $\alpha$ and $\beta$ orbits in our previous SdH study and the Berry phase $\pi$ (0) was confirmed for $\alpha$ ($\beta$) orbits.
In the present study, we perform new SdH measurements up to higher temperatures, aiming to extract the temperature dependence of the $\alpha$ and $\beta$ frequencies and to compare them to the GAM model.

\section*{Results}
\subsection*{Experimental results and analysis}
Figure 1 shows the ($B$-up) magnetoresistivity, second field derivative, and Fourier transform of $\mathrm{d}^2 \rho / \mathrm{d} B^2$ vs $1/B$ for three samples \#0618, \#0811, and \#1012.
(At each set temperature, a $B$-up and a $B$-down sweep were performed. See Methods.)
SdH oscillations are visible in all the samples.
The Fourier transform of the second derivative in $1/B$ shows the two peaks corresponding to the $\alpha$ and $\beta$ frequencies.
The insets show the temperature dependences of the oscillation amplitudes.
The amplitude error bars are based on the background amplitude observed in the spectra at higher frequencies than $\alpha$ and $\beta$, while the temperature error is assumed to be 5\%.
We estimate effective masses, as shown in Table I, by orthogonal distance regression fitting to the temperature reduction factor $R_{\mathrm{T},1}$ (solid lines).
The associated errors are calculated from the chi-square, which takes account of the amplitude and temperature errors (i.e., those errors are used as standard deviations of measurements in computing the chi-square).
The obtained values are mutually consistent and also consistent with our previous result \cite{Terashima18PRX}.

In order to precisely determine the frequencies at each temperature, we fit experimental data $\rho_{\mathrm{exp}}$ using Eq.~\ref{LK}.
The Fourier spectra (Fig. 1) indicate that harmonic contents ($r > 1$) are not appreciable.
However, since the second-harmonic frequency of $\alpha$ is close to $\beta$ (i.e., $2F_{\alpha} \approx F_{\beta}$) and hence may affect the fitting of $\beta$, we include $2\alpha$ in fitting low-temperature data (it is omitted at high temperatures as described below).
Thus, the fit function is expressed as  $\rho_{\mathrm{fit}} = \rho_0 (1+\Delta \rho_{\alpha}/\rho_0 + \Delta \rho_{2\alpha}/\rho_0 +\Delta \rho_{\beta}/\rho_0)$, where $\rho_0$ is given by a second-order polynomial.
As determined in our previous study \cite{Terashima18PRX}, $R_{\mathrm{s},1}$ is positive and $\phi_{\mathrm{D}}$ = 0 for both frequencies, while $\phi_{\mathrm{B}}$ = $\pi$ and 0 for $\alpha$ and $\beta$, respectively.
The value of the spin-splitting factor $S$ affects the ratio of the second-harmonic to fundamental amplitudes, and we fix $S$ for the $\alpha$ frequency at 0.321 and that for $\beta$ at 1.61 according to \cite{Terashima18PRX}.

We explain our fitting procedure with Fig. 2, which shows fitting results at selected temperatures for sample \#0811.
We fix the effective masses as estimated above: $m^*_{\alpha}/m_{\mathrm{e}}$ = 0.43, and $m^*_{\beta}/m_{\mathrm{e}}$ = 0.89 for sample \#0811.
The remaining parameters are $F$, $T_{\mathrm{D}}$, and a proportionality factor $C$ for each of $\alpha$ and $\beta$.
We first fit the average of the $B$-up and $B$-down data at $T$ = 0.03 K above $B$ = 5 T with those free fitting parameters.
The obtained frequencies are assumed to be the zero-temperature frequencies $F_0$ (Table I).
We then fit $B$-up and $B$-down data, separately, at each set temperature with the values of $T_{\mathrm{D}}$ and $C$ fixed at those obtained at $T$ = 0.03 K (Table I):
accordingly only $F_{\alpha}$ and $F_{\beta}$ are the fitting parameters at each temperature.
Figure 2\textbf{a} shows the fitting result for the $B$-up data at $T$ = 0.03 K.
The fit curve (black dashed) reproduces the experimental one (red) excellently.
Decomposition into components (lower curves) indicates that the $\alpha$ component $\Delta\rho_{\alpha}$ (orange) dominates the oscillatory part $\rho_{\mathrm{exp}}-\rho_0$ (red) and that the residue $\rho_{\mathrm{exp}}-\rho_0-\Delta\rho_{\alpha}$ (black dotted) is already rather small.
It also shows a phase difference of about $\pi$ between the $2\alpha$ (pink) and $\beta$ (purple) components in the measured field range.
Figure 2\textbf{b} shows that the dominance of $\alpha$ is further strengthened at $T$ = 1.82 K:
the oscillatory part (red) shown in the lower right part is almost perfectly described by the $\alpha$ component (orange).
We therefore omit 2$\alpha$ and $\beta$ at and above $T$ = 3.0 K (the maximum measurement temperature for this sample is 9 K).
Figure 2\textbf{c} shows that the oscillation at $T$ = 3.0 K is perfectly described by the $\alpha$ component alone.
From these fitting results, we obtain $\Delta F_{\alpha (\beta)} = F_{\alpha (\beta)} - F_0$ at measured temperatures.

We now like to estimate the errors in the frequencies caused by the errors in the effective masses.
We therefore repeat the above fitting procedure but with the effective-mass values increased by the amount of the error: namely, $m^*_{\alpha}/m_{\mathrm{e}}$ = 0.45, and $m^*_{\beta}/m_{\mathrm{e}}$ = 0.92.
We assume the errors in the temperature-dependent frequencies are equal to the differences between the frequencies obtained in the two fitting routines.
The errors in the Dingle temperatures are also estimated by comparing results of the two routines (Table I).

Data for sample \#0618 and \#1012 are analyzed in the same way, and examples of the fitting results are shown in Fig. 3.
For sample \#1012, the 2$\alpha$ and $\beta$ components are omitted at and above $T$ = 4.0 K (the maximum measurement temperature for sample \#1012 is 7 K, while that for \#0618 is 1.8 K).
Figure 4 shows the obtained temperature shifts of the frequencies for the three samples.
Note that there are two data points at each set temperature corresponding to $B$-up and $B$-down sweeps, though they are almost indistinguishable at most temperatures.

\section*{Discussion}

Before discussing Fig. 4, we review theoretical frequency shifts expected from the GAM model \cite{Guo21NatCommun}.
Following GAM \textit{et al.}, we introduce a parameter $\Theta$ and rewrite Eq.~\ref{GAM} as follows:
\begin{equation}\label{GAM2}
F(T) = F_0 -\Theta \left(\frac{\pi k_{\mathrm{B}}}{\mu_{\mathrm{B}}}\right)^2 \frac{T^2 (\mu^*)^2}{F_0},
\end{equation}
where $\mu_{\mathrm{B}}$ is the Bohr magneton.
GAM \textit{et al.} noted that $\Theta$ is the sum of the topological part $\Theta^{\mathrm{T}}$ = 1/16 and a Sommerfeld part $\Theta^{\mathrm{S}}$ due to the temperature dependence of the chemical potential $\zeta$ (Sommerfeld correction).
The chemical potential shift is described by
\begin{equation}\label{ChemPot}
\zeta(T) = E_{\mathrm{F}} - \frac{1}{6} (\pi k_{\mathrm{B}} T)^2 \frac{D^{\prime}}{D},
\end{equation}
where $D$ and $D^{\prime}$ are the zero-field density of states and its energy derivative, respectively.
The corresponding frequency shift can be calculated by $\Delta F = (+/-)(m^*/(e \hbar)) \Delta \zeta$ (electron/hole).
When evaluating Eq.~\ref{ChemPot}, all existing Fermi pockets have to be included.
GAM \textit{et al.} considered two extreme cases \cite{Guo21NatCommun}:
One is Cd$_3$As$_2$, which has only small Dirac pockets.
$D$ is solely determined by those pockets, which gives $\Theta^{\mathrm{S}}$ = 1/24.
The other is a tiny Dirac pocket in LaRhIn$_5$, which coexists with much larger Schr\"odinger ones. 
In this case, $D^{\prime}/D$ is dominated by the latter pockets with an effective Fermi energy much larger than measurement temperatures, and hence $\Delta \zeta$ is negligibly small, i.e., $\Theta^{\mathrm{S}}$ = 0. 
The experimental frequency shifts in the two compounds were excellent agreement with those expected from the sum of $\Theta^{\mathrm{T}}$ and $\Theta^{\mathrm{S}}$.

The present case is more general.
We have to consider the $\alpha$ Dirac and $\beta$ Sch\"odinger cylinders at the same time:
$D = 2D_{\alpha}+D_{\beta}$ since there are two $\alpha$ cylinders in the Brillouin zone.
We make a two-dimensional approximation.
Then, the density of states is proportional to the effective mass: $D_{\beta} = rD_{\alpha}$ with $r=m_{\beta}^*/m_{\alpha}^*$.
Further, since $m^* = |E|/v^2$ for Dirac electrons, $D_{\alpha}^{\prime}/D_{\alpha}=1/E_{\alpha}$, where $E_{\alpha}$ is the Fermi energy of the $\alpha$ pocket measured from the Dirac point.
On the other hand, $m^*$ is constant for parabolic bands and hence $D_{\beta}^{\prime}$ = 0.
Accordingly, $D^{\prime}/D=2D_{\alpha}^{\prime}/(2D_{\alpha}+D_{\beta}) =   2/((2+r)E_{\alpha})$.
By noting $F_{\alpha} =  m_{\alpha}^* E_{\alpha} / (2 e \hbar)$ and that $\mu_{\mathrm{B}} = e\hbar /(2m_{\mathrm{e}})$, we obtain $\Theta^{\mathrm{S}}_{\alpha}=(1/(2+r))(1/24)$.
On the other hand, $F_{\beta} =  m_{\beta}^* E_{\beta} / (e \hbar)$, where $E_{\beta}$ ($>0$) is the Fermi energy of the $\beta$ pocket measured from the top of the hole band.
Then, we obtain $\Theta^{\mathrm{S}}_{\beta}=-(R/(2+r))(1/12)$, where $R=E_{\beta}/E_{\alpha}$.
Using experimental values in Table I, we obtain $\Theta_{\alpha}=\Theta^{\mathrm{S}}_{\alpha}+\Theta^{\mathrm{T}}_{\alpha}$ = 0.0719(6), 0.0727(5), and 0.072(2), and $\Theta_{\beta} = \Theta^{\mathrm{S}}_{\beta}$ = -0.010(2), -0.012(2), and -0.011(4) for sample \#0618, \#0811, and \#1012, respectively.
We show in Fig. 4 the theoretical frequency shifts calculated for sample \#0811 (the shifts calculated for the other samples are almost indistinguishable from the plotted ones).

Our main results shown in Fig. 4 are qualitatively consistent with the GAM model in that the $\alpha$ frequency arising from the Dirac pockets shows a clear negative shift, while the trivial $\beta$ frequency does not.
Further, at high temperatures where the amplitude of the $\beta$ oscillation is negligibly small, the experimental $\alpha$-frequency shifts show linear dependence on $T^2 (\mu^*)^2 / F_0$.
Linear fits to \#0811 and \#1012 data points for $T \geqslant 5$ K (broken lines) give $\Theta$ = 0.09(2) and 0.08(2), respectively, where the errors are estimated from the fitting error, which takes account of the errors in the frequency and temperature, and the systematic error due to the error in the effective mass.
The obtained values agree with the theoretical values estimated above within experimental accuracy.
This observation supports the quantitative accuracy of the GAM model.

On the other hand, at low temperatures where the $\beta$ oscillation is not negligible, the experimental frequency shits largely deviate from the theoretical lines for both the $\alpha$ and $\beta$ frequencies and exhibit appreciable sample dependence.
We suspect that the two frequencies could not be determined accurately at those low temperatures.
As noted above, the phase difference between the $2\alpha$ (pink) and $\beta$ (purple) oscillations is approximately $\pi$ in the measured field range (Figs. 2 and 3).
This destructive interference most likely causes ambiguity in the fitting procedure and hence results in erroneous estimates of the $\alpha$ and $\beta$ frequencies.
In the present measurements, the observed oscillation periods are only a few or less.
In order to accurately determine the two frequencies at the same time, it is necessary to observe more oscillation periods in a wider $1/B$ window so that the phase difference between 2$\alpha$ and $\beta$ deviates from $\sim\pi$.
We mention in passing that, although the reasonable values of $\Theta$ (i.e., slopes) were estimated from the $\alpha$-frequency shifts at high temperatures as described above, the absolute values of the shits themselves are larger than theoretically expected.
This can also be ascribed to that the frequencies could not be determined very accurately at low temperatures (i.e., the estimation of $F_0$ is inaccurate).

To summarize, we performed SdH measurements on CaFeAsF from $T$ = 0.03 K to a high temperature ($T$ = 9 K for sample \#0811 and 7 K for sample \#1012).
The SdH frequency of the $\alpha$ Dirac cylinder showed a $T^2$ negative shift at sufficiently high temperatures where the $\beta$ oscillation is negligible, and the rate of the frequency shift is consistent with the topological frequency shift predicted by GAM \textit{et al.} \cite{Guo21NatCommun}.
At low temperatures where the $\beta$ oscillation is not negligible, the temperature shifts of the two frequencies $\alpha$ and $\beta$ deviate from those expected from the GAM model.
We ascribed this to that the two frequency could not be determined accurately enough because of unfavorable interference between the 2$\alpha$ and $\beta$ oscillations.
If we are to compare frequency shifts at low temperatures to the GAM model, further measurements on better samples are necessary to observe more oscillation periods and hence to determine the two frequencies more accurately.
Nonetheless, the present work as a whole confirms that the GAM method can be used to identify Dirac/Weyl pockets provided that quantum-oscillation frequencies can be determined with sufficient accuracy.
We further mention this:
GAM \textit{et al}  \cite{Guo21NatCommun} demonstrated the validity of the GAM method by measurements on Cd$_3$As$_2$, which has only a small Dirac pocket, and LaRhIn$_5$, in which a tiny Dirac pocket coexists with far larger Fermi pockets.
In CaFeAsF, similarly-sized small electron and hole pockets coexist.
This situation is more relevant to many topological semimetals.
Thus the present work constitutes an experimental demonstration of a wider applicability of the GAM method.
Finally, the GAM method is complementary to the phase analysis of quantum oscillation, which has its own shortcoming that the phase is affected not only by the Berry phase but also by other factors such as the spin reduction factor.

\section*{Methods}
\subsection*{Samples and measurements}
CaFeAsF single crystals \#0618, \#0811, and \#1012 were prepared in Shanghai by a CaAs self-flux method as described in \cite{Ma15SST}.
The resistivity was measured along the $c$ axis, and the magnetic field up to 17.5 T was applied along the $c$ axis.
The electrical contacts were spot-welded and reinforced with silver conducting paint.
Since both $\alpha$ and $\beta$ pockets are cylinders extended along the $c$ axis, it is reasonable to assume that off-diagonal terms of $\rho$, $\rho_{ca}$ and $\rho_{cb}$, are negligible for this geometry and hence that $\sigma_{cc} = \rho_{cc}^{-1}$, which allows the use of Eq.~\ref{LK}. 
A dilution refrigerator with a base temperature of 0.03 K was used, and the maximum measurement temperature was 1.8 , 9, and 7 K for sample \#0618, \#0811, and \#1012, respectively (sample \#0618 was measured first, and we did not notice that higher temperatures were necessary).
At each set temperature, the magnetic field was ramped up to 17.5 T and then down to zero, and thus two magnetoresistivity curves ($B$-up and $B$-down) were obtained.
No hysteresis was observed and $B$-up and $B$-down curves agreed well at all temperatures.

\section*{Data availability}
The data that support the findings of this study are available from the corresponding authors upon reasonable request.

\begin{acknowledgments}
This work was supported in Japan by JSPS KAKENHI Grant Number 17K05556.
This work was supported in China by the Youth Innovation Promotion Association of the Chinese Academy of Sciences (No. 2015187).
The authors thank Yuki Fuseya for valuable discussion.
\end{acknowledgments}

\section*{Competing interests}
The Authors declare no Competing Financial or Non-Financial Interests.

\section*{Author contributions}
TT planned the project and wrote the manuscript.  TW and GM prepared the samples.  TT and SU performed the measurements and analyzed the data.

\newpage
\section*{References}

\newpage

\begin{figure*}
\includegraphics[width=16cm]{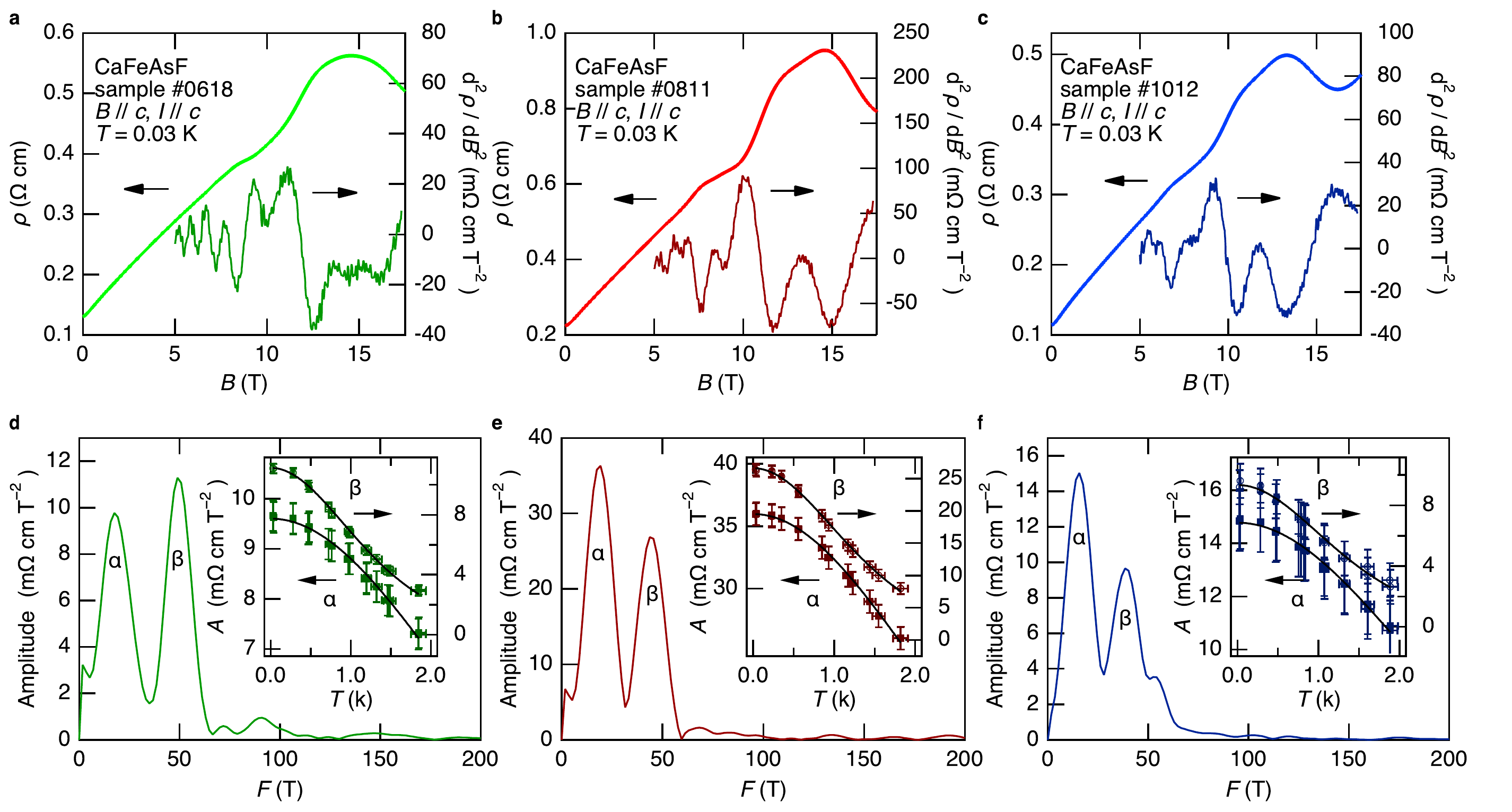}
\caption{Shubnikov--de Haas oscillation in CaFeAsF.  \textbf{a}, \textbf{b}, and \textbf{c} Magnetoresistivity, its second field derivative, and \textbf{d}, \textbf{e}, and \textbf{f} Fourier transform of $\mathrm{d}^2 \rho / \mathrm{d} B^2$ vs $1/B$ for samples \#0618, \#0811, and \#1012, respectively.  Since $B$-up and $B$-down data are virtually indistinguishable, only the $B$-up data are shown.  (insets) Temperature dependence of $\alpha$ and $\beta$ oscillation amplitudes.  The amplitude error bars are based on the noise floor of the Fourier spectra, and the temperature error bars are assumed to be 5\%.  The solid lines are fits to $R_{\mathrm{T},1}$.}
\end{figure*}

\newpage

\begin{figure*}
\includegraphics[width=10cm]{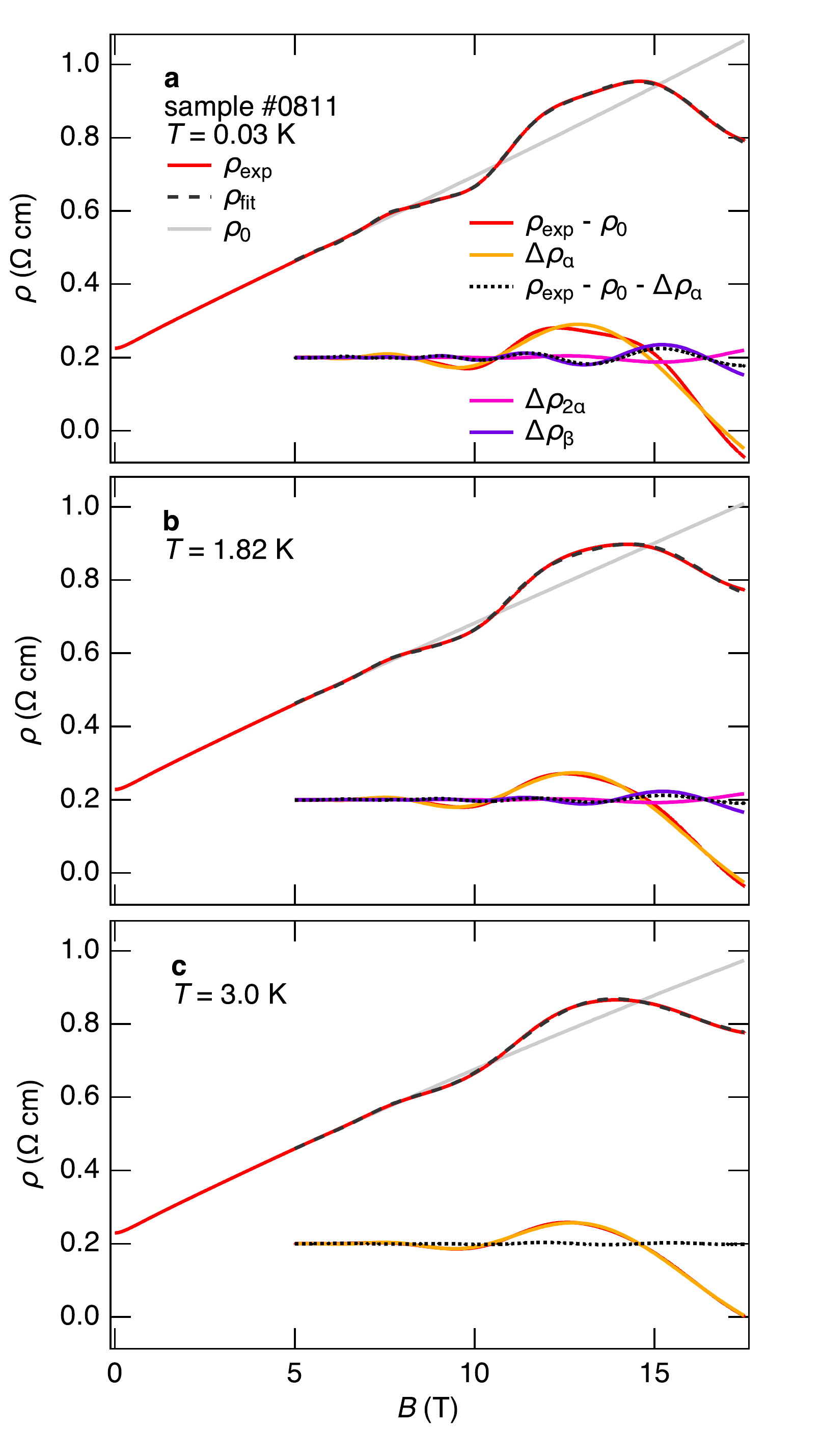}
\caption{Lifshitz-Kosevich fits to the magnetoresistivity in sample \#0811 for selected temperatures.  Only $B$-up data are shown.  The total resistivity $\rho_{\mathrm{exp}}$ (red) and fit $\rho_{\mathrm{fit}}$ (dashed black) are shown with the estimated smooth background $\rho_0$ (grey).  The oscillatory part $\rho_{\mathrm{exp}}-\rho_0$ (red) is shown in a lower right part of each figure with an upward shift of 0.2 $\Omega$cm.  It is decomposed into the $\alpha$ (orange), 2$\alpha$ (pink), and $\beta$ (purple) components at $T$ = 0.03 K (\textbf{a}) and 1.82 K (\textbf{b}), while only the $\alpha$ component is enough to reproduce the oscillatory part at $T$ = 3.0 K (\textbf{c}) (the oscillatory part and $\alpha$ component are almost indistinguishable in the figure).  Note that dotted black lines $\rho_{\mathrm{exp}}-\rho_0-\Delta\rho_{\alpha}$ are residues left after only the $\alpha$ component is subtracted from the oscillatory part.}
\end{figure*}

\newpage

\begin{figure*}
\includegraphics[width=10cm]{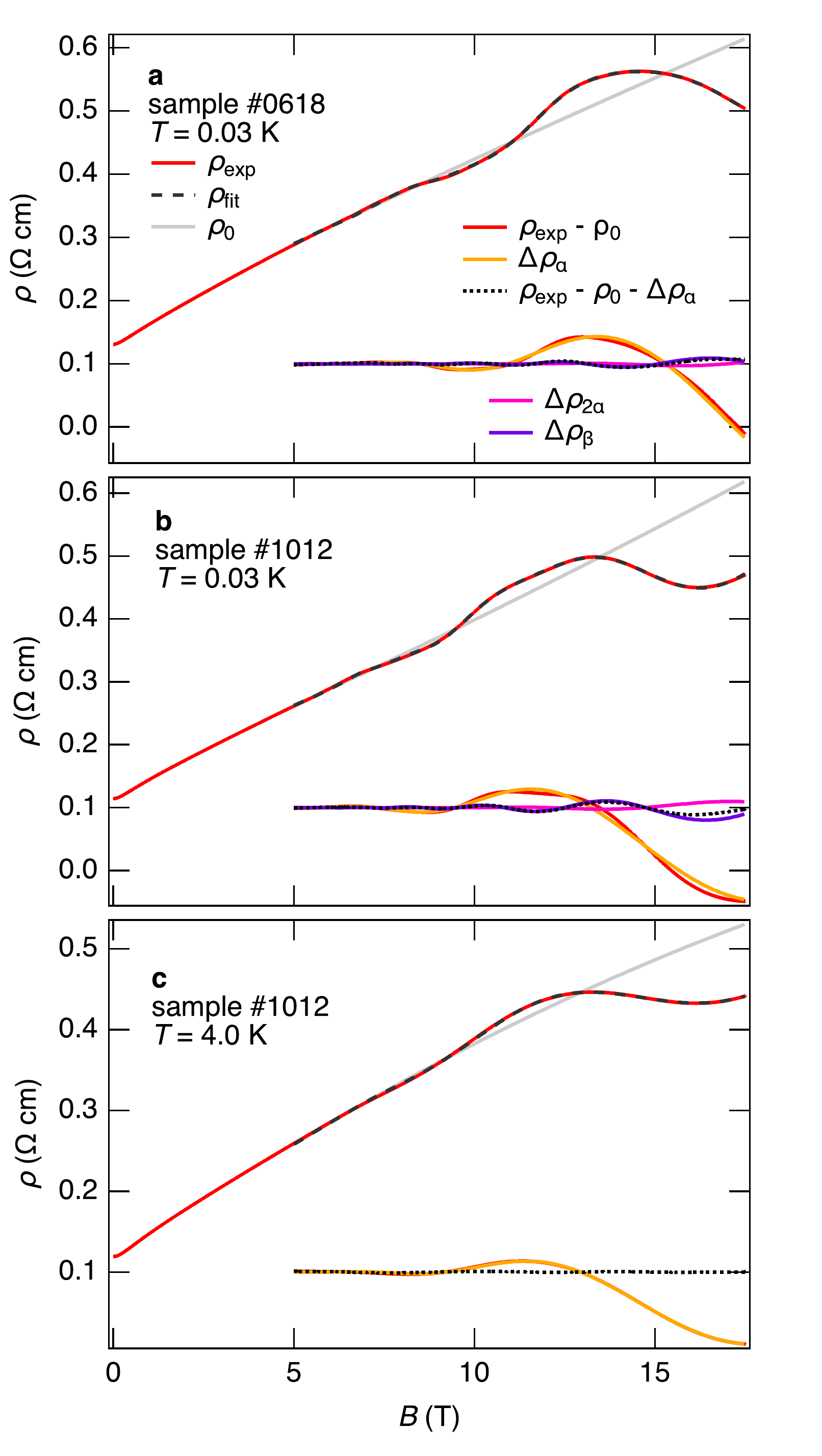}
\caption{Lifshitz-Kosevich fits to the magnetoresistivity in samples \#0618 and \#1012 for selected temperatures.  Only $B$-up data are shown.  The total resistivity $\rho_{\mathrm{exp}}$ (red) and fit $\rho_{\mathrm{fit}}$ (dashed black) are shown with the estimated smooth background $\rho_0$ (grey).  The oscillatory part $\rho_{\mathrm{exp}}-\rho_0$ (red) is shown in a lower right part of each figure with an upward shift of 0.1 $\Omega$cm.  It is decomposed into the $\alpha$ (orange), 2$\alpha$ (pink), and $\beta$ (purple) components at $T$ = 0.03 K (\textbf{a} and \textbf{b}), while only the $\alpha$ component is enough to reproduce the oscillatory part in \#1012 at $T$ = 4.0 K (\textbf{c}) (the oscillatory part and $\alpha$ component are almost indistinguishable in the figure).  Note that dotted black lines $\rho_{\mathrm{exp}}-\rho_0-\Delta\rho_{\alpha}$ are residues left after only the $\alpha$ component is subtracted from the oscillatory part.}
\end{figure*}

\newpage

\begin{figure*}
\includegraphics[width=16cm]{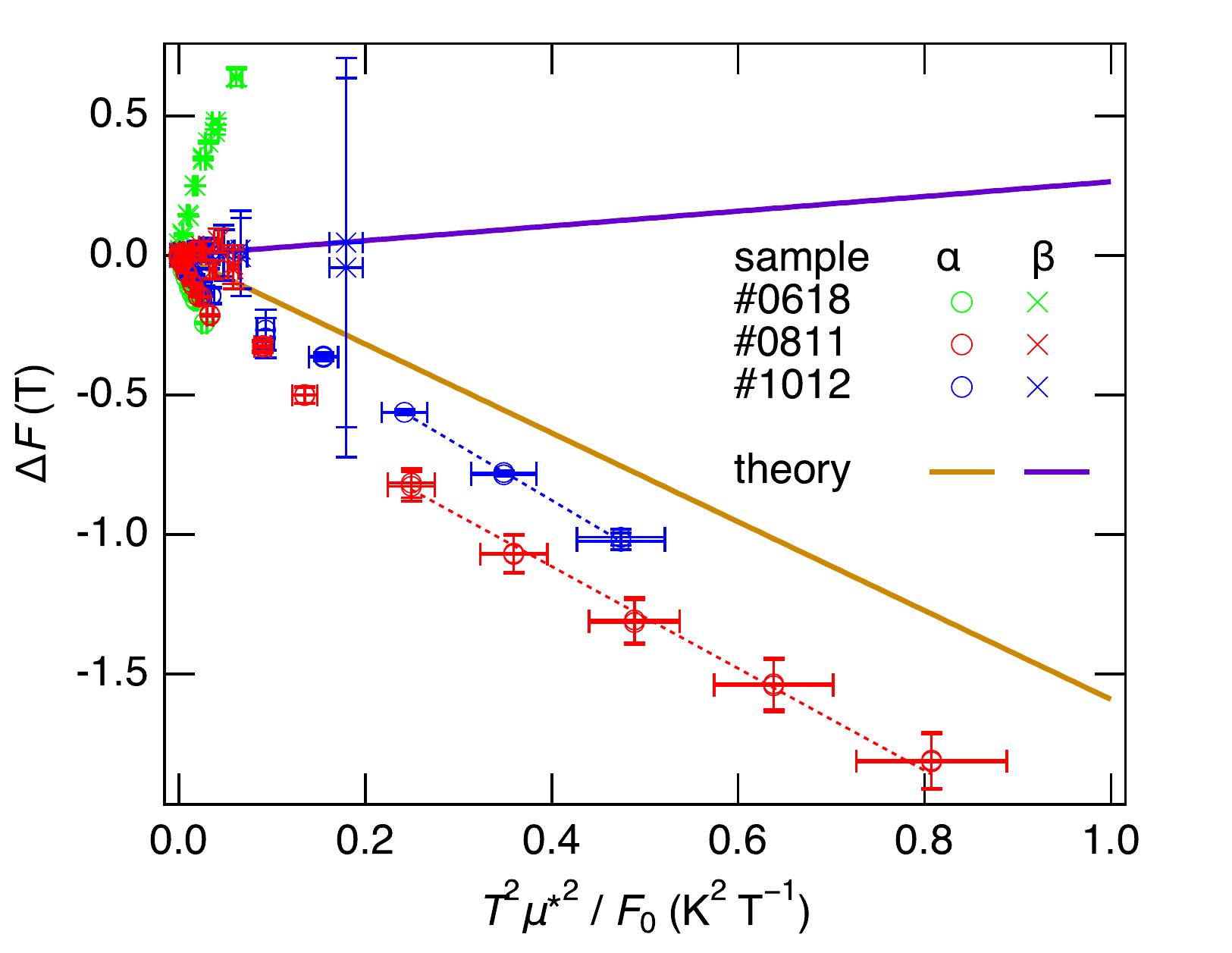}
\caption{Experimental (marks) and theoretical (solid lines) frequency shifts in CaFeAsF.  The estimation of the vertical error bars is explained in the main text.  The horizontal error bars are based on the 5\% temperature error (the systematic error due to the error in $\mu^*$ is not shown).  The broken lines are linear fits to data points of \#0811 and \#1012 for $T\geqslant 5$ K}
\end{figure*}

\begin{table*}
\caption{\label{Tab} Fermi-surface parameters.  The zero-temperature frequencies ($F_0$) are estimated from Lifshitz-Kosevich fits to base-temperature data, and the associated errors are numerical fitting errors.  The effective masses ($m^*$) are estimated from the temperature dependence of Fourier amplitudes, and the associated errors take into account the amplitude and temperature errors in experiment.  The Dingle temperatures ($T_{\mathrm{D}}$) are estimated from Lifshitz-Kosevich fits to base-temperature data, and the associated errors take into account the errors in the effective masses.  See text for details of the error estimation.  The Fermi energies ($E_{\alpha}$ and $E_{\beta}$) are derived from $F_0$ and $m^*$, using the linear dispersion relation ($E_{\mathrm{F}} = \hbar v_{\mathrm{F}} k_{\mathrm{F}}$) for the $\alpha$ cylinder and a quadratic one ($E_{\mathrm{F}} = \hbar^2 k_{\mathrm{F}}^2 / (2m^*)$) for $\beta$.}
\begin{ruledtabular}
\begin{tabular}{ccccccccc}
 & \multicolumn{4}{c}{$\alpha$} & \multicolumn{4}{c}{$\beta$}\\
sample & $F_0$ (T) & $m^*/m_{\mathrm{e}}$ & $T_{\mathrm{D}}$ (K) & $E_{\alpha}$ (meV) & $F_0$ (T) & $m^*/m_{\mathrm{e}}$ & $T_{\mathrm{D}}$ (K) & $E_{\beta}$ (meV) \\
\hline
\#0618 & 19.121(2) & 0.39(3) & 7.6(6) & 11.4(9) & 49.044(8) & 0.94(3) & 2.53(8) & 6.0(2)\\
\#0811 & 18.542(2) & 0.43(2) & 4.8(3) & 10.0(5) & 45.222(5) & 0.89(3) & 2.45(8) & 5.9(2)\\
\#1012 & 16.527(2) & 0.40(5) & 5.9(7) & 10(2) & 40.522(3) & 0.87(8) & 2.3(2) & 5.4(5)\\
\end{tabular}
\end{ruledtabular}
\end{table*}

\end{document}